\documentclass{eptcs}
\providecommand{\event}{UITP 2016} 

\usepackage[T1]{fontenc}
\usepackage[utf8]{inputenc}
\usepackage[english]{babel}

\usepackage{mdframed}
\usepackage{scrextend}

\usepackage[final]{fixme}
\FXRegisterAuthor{pj}{anpj}{PJ}
\FXRegisterAuthor{eg}{aneg}{EG}
\FXRegisterAuthor{bp}{anbp}{BP}

\usepackage{xcolor}
\usepackage{graphicx}

\usepackage{listings}
\usepackage{paralist}

\definecolor{lightgray}{rgb}{.9,.9,.9}
\definecolor{darkgray}{rgb}{.4,.4,.4}
\definecolor{purple}{rgb}{0.65, 0.12, 0.82}

\lstdefinelanguage{JavaScript}{
  keywords={typeof, new, true, false, catch, function, return, null, catch, switch, var, if, in, while, do, else, case, break},
  keywordstyle=\color{blue}\bfseries,
  ndkeywords={class, export, boolean, throw, implements, import, this},
  ndkeywordstyle=\color{darkgray}\bfseries,
  identifierstyle=\color{black},
  sensitive=false,
  comment=[l]{//},
  morecomment=[s]{/*}{*/},
  commentstyle=\color{purple}\ttfamily,
  stringstyle=\color{red}\ttfamily,
  morestring=[b]',
  morestring=[b]"
}

\lstnewenvironment{jslisting}%
  {\lstset{language=JavaScript}}
  {}

\lstnewenvironment{ocaml}%
  {\lstset{language=caml}}
  {}

\usepackage{xspace}
\newcommand{\jscoq}{jsCoq\xspace}

\newcommand{\coq}{Coq\xspace}
\newcommand{\coqdoc}{CoqDoc\xspace}
\newcommand{\jsoo}{js\_of\_ocaml\xspace}

\title{\jscoq: Towards Hybrid Theorem Proving Interfaces}
\author{Emilio Jesús Gallego Arias
\institute{MINES ParisTech\\PSL Research University, France}
\email{e@x80.org}
\and
Benoît Pin
\institute{MINES ParisTech\\PSL Research University, France}
\email{benoit.pin@mines-paristech.fr}
\and
Pierre Jouvelot
\institute{MINES ParisTech\\PSL Research University, France}
\email{pierre.jouvelot@mines-paristech.fr}
}
\def\titlerunning{\jscoq: Towards a Hybrid Theorem Proving Interface
  for Coq}
\def\authorrunning{E. J. Gallego Arias, B. Pin, \& P. Jouvelot}

\begin{document}
\maketitle

\begin{abstract}
  We describe \jscoq, a new platform and user environment for the Coq
  interactive proof assistant. The \jscoq system targets the
  HTML5--ECMAScript 2015 specification, and it is typically run inside a
  standards-compliant browser, without the need of external servers or
  services.

  Targeting educational use, \jscoq allows the user to start
  interaction with proof scripts right away, thanks to its
  self-contained nature. Indeed, a full \coq environment is
  packed along the proof scripts, easing distribution and
  installation. Starting to use \jscoq is as easy as clicking on a link.
  The current release ships more than 10 popular \coq libraries, and
  supports popular books such as \emph{Software Foundations} or
  \emph{Certified Programming with Dependent Types}.

  The new target platform has opened up new interaction and display
  possibilities. It has also fostered the development of some new
  Coq-related technology. In particular, we have implemented a new
  serialization-based protocol for interaction with the proof
  assistant, as well as a new package format for library distribution.
\end{abstract}

\section{Introduction}

Interactive Theorem Proving (ITP) relies on mutual human-machine
feedback to build proofs by refinement. Typically, the user first
requests the proof assistant to validate or guess some proof
step. Depending on the output of the tool, she will continue the proof
or correct the last step.

Teaching ITP is usually practice-led; students are encouraged to
interact with the tools from the start, simultaneously discovering
their way into the particularities of the implementations and the
logical theories behind them.
Popular ITP teaching material, such
as~\cite{Pierce:SF,chlipalacpdt2011}, is written in a \emph{literate
  programming style}, pioneered by Donald Knuth. In literate
programming, source code and comments meld to form a
coherent, self-documenting book and program. An additional design objective
in the context of proof assistants is that the user \emph{must be able
  to interact with the book}.
The same principles apply to documentation, which may be hard to
understand at first without the ability to run and play with examples.

Narrowing down our attention to the \coq proof
assistant~\cite{Coq:manual}, most teaching material is consumed by students with the
help of specific Integrated Development Environments (IDE). Popular
options are CoqIde or
ProofGeneral~\cite{DBLP:conf/tacas/Aspinall00,CompanyCoq2016}, but
alternatives exist. IDEs usually contain or are based on text-editing
programs to provide specialized support for proof development.
This approach works well in practice; however, IDEs are mostly focused
on proof development and provide varying degrees of support for
document-like features. For instance, not all IDEs support images,
advanced formatting, hyperlinks, or dynamic content creation. This
hinders readability and it is not uncommon to see students with the
same document doubly opened in a document reader and the IDE, as it is
difficult to make sense of the document without the possibility of
interaction. Additionally, the specialized nature of these tools makes installation
sometimes heavy, especially if third-party add-ons are involved, posing a
barrier to the casual learner.

The \jscoq system intends to tackle both of these problems (readability and installation) from a document point
of view: instead of equipping an IDE with document-like features, we
extend documents to provide IDE-like capabilities. Our approach is to
profit from modern browsers and web standards, embedding \coq as an
application inside the browser. In this setting, \coq scripts are
plain HTML documents, and a full \coq instance is run locally in the
browser for every document. A document manager --- written in
JavaScript --- manages the communication between the browser and
the \coq instance.
This setup provides a working \coq environment in a transparent way
for the user. She will just click on a link, and have an interactive \coq
document ready without any other special action or installation, even
if the script depends on exotic libraries or add-ons. The document and
proof assistant are distributed together and maintained in sync;
moreover, the standards-based source code of \jscoq should ensure a good amount of forward
compatibility.

There are many goals to consider when porting a system of the sheer
size of \coq (currently more than 200.000 lines of source code), which we address here.
\begin{labeling}{Maintenability}
\item[Completeness] The full \coq system should run in the browser platform, since non-trivial \coq courses require large helper libraries, from real numbers to complex decision procedures. 
\item[Relevance] We intend to support from the very start existing teaching material;
\item[Performance] To be practical, \jscoq should be able to handle large real-world developments in a usable way, and it would be
disappointing to find in \jscoq a crippled \coq system.
\item[Usability] We rely for the front end on modern web technologies, which carry a certain
\emph{coolness} factor that may motivate the students to use and contribute to the tool.
\item[Maintainability] The architecture of \jscoq  should be
such that keeping up with upstream\footnote{Upstream tools and technologies are those on which a given system relies, such as \coq in our case.} changes in \coq is low-effort, allowing it to run over an unpatched, up-to-date \coq version.
\end{labeling}

\noindent 
In fact, we believe we started the project at the exact moment when
achieving all these goals became possible, as it is the combination of
very recent improvements in the \coq API, OCaml-to-JavaScript
technology, and web standardization that made the project successful.

The \jscoq development is managed in an open-source way. Our
\href{https://github.com/ejgallego/jscoq/}{project
  page}~\cite{jscoq-github} provides information on source code, user
mailing lists, builds, settings, packages, compatibility, and
installation instructions. Note that the intrinsic nature of this
project will unavoidably make obsolete some of the information
contained in this paper, we recommend users to check our project page
for up-to-date information.

The structure of the paper is as follows. 
We start with a brief introduction to our Web-Based Hybrid Document
Model in
Section~\ref{sec:hybr-docum-model}. Section~\ref{sec:description-system}
gives an overview of the main components and architecture of \jscoq. We report on some preliminary practical use of \jscoq for education purposes in Section~\ref{sec:practical-validation}.
We discuss related work in Section~\ref{sec:related-work} and address possible future work in Section~\ref{sec:future-work}. We conclude in Section~\ref{sec:conclusion}. 

\section{The Web-Based Hybrid Document Model}
\label{sec:hybr-docum-model}

We use the term \emph{hybrid document}\footnote{or \emph{rich
    document}, \emph{interactive document}, or any of the many names
  used in the literature for this concept} to denote documents capable
of containing objects whose interpretation is given by an external
program.
A typical example is a word processor file embedding a spreadsheet or, in
our case, an HTML page containing a Coq proof script.
The hybrid-document platform of choice for \jscoq is the so-called
\emph{modern web platform} specified by the
ECMAScript\textsuperscript{\textregistered} 2015~\cite{EcmaScript:15}
standard. In it, base documents are defined using the HTML markup
language. JavaScript --- a Turing-complete, object-oriented, and
functional language --- is used to manipulate HTML documents by means
of the \emph{Document Object Model} API. 

The modern web platform enjoys wide adoption and support, with increasing
compatibility and standardization making it ever more
future-proof. Additionally, the ubiquity of JavaScript code in the web
has led to huge improvements in execution performance: browsers can
run now applications of size and complexity thought unmanageable some
years ago.  The process known as \emph{transpiling}\footnote{the
  process of compiling a language to another, i.e., in this case, to
  JavaScript} thus allows developers to port native applications to
JavaScript without a full rewrite. The number of libraries available
is staggering; a document can easily gain advanced 3D mathematical
plotting~\cite{mathbox}, editing capabilities~\cite{codemirror}, or
even \LaTeX\ support~\cite{mathjax}, with just a few lines of code.

There are many interesting examples of web-based hybrid documents.
Jupyter\footnote{previously known as IPhyton} allows to create and
share documents that contain live code, equations, visualizations, and
explanatory text; Eloquent JavaScript~\cite{eloquent-js} is an
interactive book where all the examples can be run and edited online.

\paragraph{Web-Based Coq Documents}

For \jscoq documents, we have chosen a simple solution: the user
provides an arbitrary HTML document, tagging \coq code as
appropriate. Generally, this means that non-editable code will be
wrapped in \lstinline!code! tags, whereas editable content will be
placed inside \lstinline!textarea! elements. At the end of the
document, the user asks for \jscoq to be loaded, providing an ordered
list of identifiers of the elements that \jscoq will understand as the
\coq document:
\begin{jslisting}
  <script src="js/jscoq-loader.js" type="text/javascript"></script>
  <script type="text/javascript">
    loadJsCoq('./').then( () => new CoqManager (list_of_ids, [options]));
  </script>
\end{jslisting}
The \lstinline!CoqManager! will scan the corresponding elements,
initializing the pertinent editor components, etc. Finally, a
CoqIDE-style control panel will be attached to the document,
containing the goals window, logs, toolbars, etc. The results can be
seen in Figure~\ref{fig:jscoq-demo}. For a live demonstration,
\href{https://x80.org/rhino-coq/}{\jscoq landing page} provides a
simple example; the best way to get a feeling of the current system is
to try it out. We list here more examples that we developed:
\begin{figure}[t]
  \centering
  \begin{mdframed}
  \includegraphics[width=\textwidth]{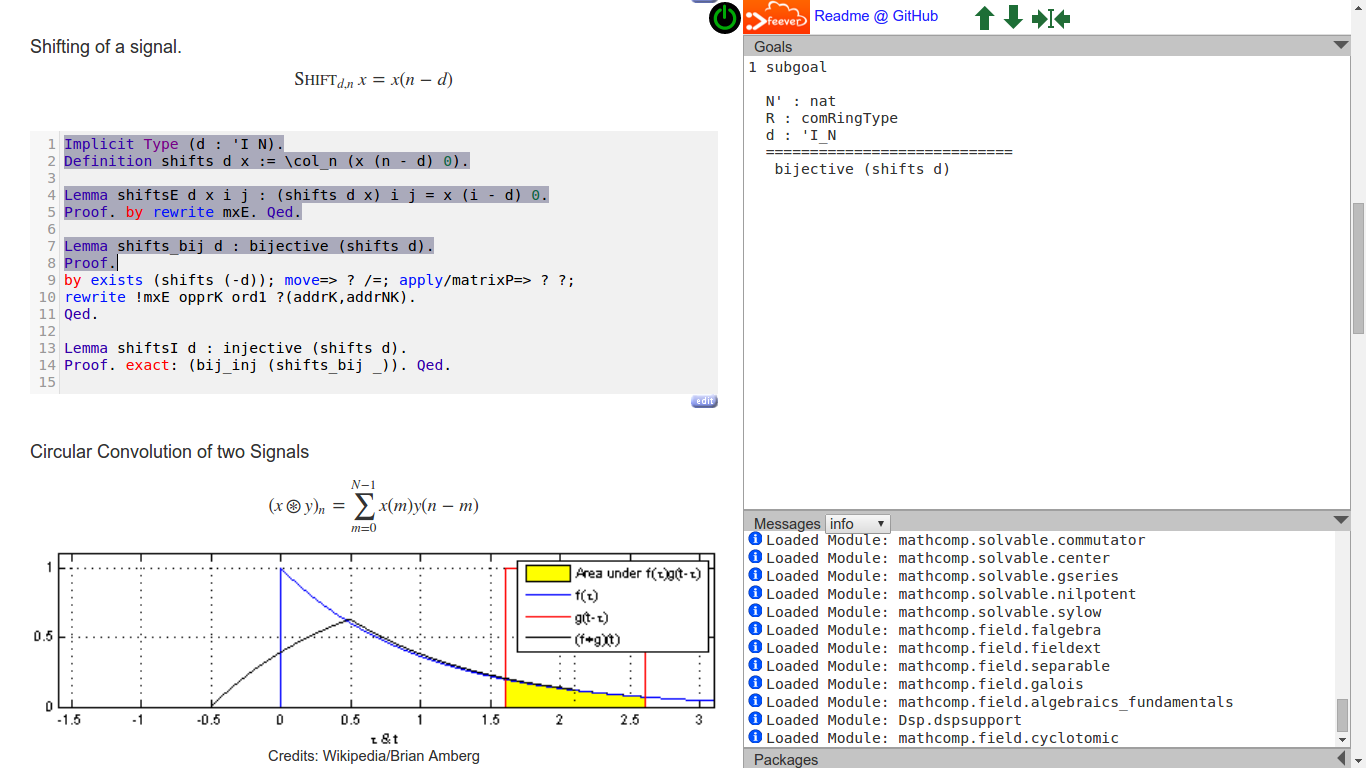}
  \end{mdframed}
  \caption{jsCoq 0.6 running a DFT formalization}
  \label{fig:jscoq-demo}
\end{figure}
%
\begin{itemize}
\item \href{https://x80.org/rhino-coq/examples/dft.html}{DFT}, a small
  development of the theory of the Fourier Transform,
  following~\cite{jos:mdft};
\item
  \href{https://x80.org/rhino-coq/examples/mtac_tutorial.html}{Mtac},
  the Mtac~\cite{DBLP:conf/icfp/ZilianiDKNV13} tutorial;
\item \href{https://x80.org/rhino-coq/examples/Stlc.html}{STLC}, the
  "Simply Typed Lambda Calculus" chapter from~\cite{Pierce:SF};
\item
  \href{https://x80.org/rhino-coq/examples/Cpdt.StackMachine.html}{StackMachine},
  the first chapter of~\cite{chlipalacpdt2011}.
\end{itemize}

\paragraph{Web Runtime Platforms}

We briefly comment on some aspects of our platform of
choice. Particular issues of interest are compatibility
and future-proofness.
Regarding compatibility, we try to be standards-compliant, and \jscoq
is known to work with different versions of Microsoft Edge, Mozilla
Firefox, and Safari. However, we recommend using Google Chrome, as it
is the browser we internally use for testing. While maybe an unpopular
decision, we lack the resources to perform proper multi-browser
testing; however we would be very happy to include fixes that help
support in other browsers, but so far those have not been
necessary. We should also note that we have run \jscoq in the Node
JS~\cite{nodejs} platform successfully. We also believe that the our
approach should be acceptably resistant to future changes in browsers
and platforms. We have taken care to restrict our development to
standard constructions that should be respected by the upstream
developers; however note that due to the technology used, \jscoq
usually requires a pretty recent browser.

\section{System Overview}
\label{sec:description-system}

The high-level architecture of \jscoq is the typical one for IDEs: the \coq
proof assistant runs in its own, separate process, whereas the user
interface runs in the browser thread. The two communicate using
message passing following the Web Worker Specification standard \cite{web-workers}. Additionally, a library manager takes care of downloading
\coq packages and registering them with the browser virtual filesystem.
\begin{description}
\item[The \coq worker]
  \coq offers an XML-based communication protocol --- used by CoqIDE
  and others --- which allows to incrementally build proof
  scripts. However, the protocol relies on Unix features that are not
  available in a browser environment, and does not provide direct
  access to \coq's internal document API which we find valuable.
  We have thus chosen to implement a simple RPC protocol on top of the
  OCaml API of \coq. This choice has saved a considerable amount of
  work on interfacing with XML objects and provided us with somewhat
  more flexibility and freedom, without interfering with the
  established XML-protocol.

  The worker relies on a new domain-specific language (DSL) for communication with
  \coq. Together, the language interpreter and definition are about
  200 lines of code. Automatic serialization from/to JSON is performed
  by the PPX mechanism of OCaml. The worker is then linked with \coq and compiled by
  \jsoo~\cite{DBLP:journals/spe/VouillonB14} to a monolithic
  JavaScript file. When instantiated, the worker will
  constitute a full \coq instance, waiting for control messages.
\item[The \coq package manager] The \coq library system is designed
  under the assumption of an underlying Unix filesystem. However,
  synchronous file I/O is not available in the browser; thus \jscoq
  needs to manage available libraries and their exact location. For
  historical reasons, the manager is currently written in OCaml and
  linked with the worker. However, there is no reason it could not
  be rewritten in JavaScript and moved out of the \coq worker.
\item[The user interface]  Written in ECMAScript, the main
  responsibility of this part is to maintain the user-side document
  model, taking care of navigation, and synchronizing it with the \coq
  worker.
\end{description}
Before describing these parts in more detail below, we would like to take a quick look at the (admittedly short) evolution of \jscoq. The first versions of \jscoq did not use a web worker thread; instead, \coq
ran in the main thread. This was a conscious design choice taken to
keep things simple and speed up the development. In this implementation,
\coq exported a synchronous \lstinline!jscoq! object with method
calls. This worked pretty well and did not require any kind of protocol,
but, in the long term, the worker solution is desirable for UI latency
purposes. Experience with that first design led to the development of the
\emph{SerAPI} protocol (see Section~\ref{sec:future-work}), which forms
the basis for the current protocol, presented below.

\subsection{\coq Protocol}
\label{sec:protocol}

The \jscoq protocol is still in evolution as detailed in
Section~\ref{sec:future-work}. We thus present a snapshot that should
provide the reader with the general ideas behind it. We recommend
consulting the actual source code to see all the low-level
details. Also, for space reasons, we omit some non-interesting calls
used to gather version information, etc.

The protocol is defined by the following DSL:
\begin{ocaml}
type jscoq_cmd =
  | Init    of string list list * string list list
  | Add     of stateid * int * string
  | Cancel  of stateid
  | Observe of stateid
  | Goals   of stateid
  | SetOpt  of bool option * string list * gvalue
  | GetOpt  of string list

type jscoq_answer =
  | Added     of stateid
  | Cancelled of stateid list
  | GoalInfo  of stateid * richpp
  | Feedback  of feedback
  | CoqOpt    of gvalue
  | CoqExn    of loc option * (stateid * stateid) option * richpp
  | JsonExn   of string
\end{ocaml}
Our message-based API is based on the State Transaction Machine
(STM) API for \coq~\cite{DBLP:conf/itp/BarrasTT15}.
The STM API was designed for asynchronous proof processing
and allows to incrementally build and edit \coq documents. Such design
has been proven invaluable in our project.
We have introduced minor differences for UI-related convenience; in
particular, we have a simpler error handling and we have replaced the
``document editing'' operation by a ``document cancellation''
one. This makes the mapping between the \coq document
model an the typical editor document model simpler.

For commands, \lstinline!Init(loadpath, init_mods)! will initialize
\coq and set up the proper loadpath and initial libraries;
\lstinline!Add(sid,eid,cmd)!  will add \lstinline!cmd! to the \coq
internal document on top of state \lstinline!sid!, with
\lstinline!eid! used to report parsing errors.
\lstinline!Cancel(sid)! will cancel a previously defined state and
their dependent states; \lstinline!Observe(sid)! will commit to the
given state. Finally, \lstinline!Goals(sid)! asks for the \coq goals
to be proved in state \lstinline!sid!.

The first answers mostly correspond to the first commands; for
instance, \lstinline!GoalInfo(sid, s)! provides a pretty-printed
string \lstinline!s! for the goal in state \lstinline!sid!.
\lstinline!Feedback! notifies the UI of changes in the \coq state
machine, such as states becoming ready. Fatal errors are signaled by
\lstinline!CoqExn(loc,(sid,fid),msg)!, with \lstinline!sid! the first
good state identifier and \lstinline!fid! the state identifier that
produced the exception.

A typical editing session will thus consist in several
\lstinline!Add! commands as the user writes a document followed by
\lstinline!Observe! when execution of a state is requested. \coq will
in turn asynchronously acknowledge parsing and execution begin and
end points, information that is used by the front end to color the sentence
appropriately. When the user edits an already submitted sentence, a
\lstinline!Cancel! event will be sent to \coq, which will in turn
respond with a list of sentences to cancel. We thus free the UI of
having to take care of sentence dependencies. \lstinline!GoalInfo!
will be performed by the UI whenever it wants to display goals. For
instance, when adding many sentences at once, \jscoq will only request
goals for the last successful one.
``Document backtracking'' can be understood in two different ways in
this protocol. If the users want to return to a previous document
state, observing that document state will move the current state to
that area. If, additionally, the user wants to change the document, a
cancel must be issued.

In addition, to allow the UI to request information about and load
available \coq packages, informing the UI back about
progress/completion, we designed a dedicated library manager control
language. It is defined as follows:
\begin{ocaml}
type lib_cmd =
  | GetInfo
  | InfoPkg of string * string list
  | LoadPkg of string * string

type lib_event =
  | LibInfo     of string * Jslib.coq_bundle
  | LibProgress of progress_info
  | LibLoaded   of string
\end{ocaml}

\subsection{Serialization}

The previous command and answer definitions are given as OCaml
datatypes, and implemented by a reasonably straightforward OCaml interpreter.
However, the browser side, and, in particular, the Web Worker API,
requires messages to be JSON objects. What is the best way to relate
our OCaml datatypes to their JSON representation?
We think that the direct manipulation of JSON objects in OCaml would have
consumed most of our available development time.  Thus, core to our
design is the use of the new PPX system for OCaml meta-programming,  which
supports the automatic generation of serializers.

In particular, we use the \texttt{ppx\_yojson}~\cite{ppx-yojson} package, which
will automatically generate serialization functions for the
definitions of our small DSL and --- more importantly --- for
core \emph{\coq data types}. Indeed, this has proven to
be a great choice, resulting in a very low overhead when tracking upstream
changes, exporting complex structures, or just experimenting.
A \coq datatype is declared serializable using pragmas, such as in the following
declaration:
\begin{ocaml}
type feedback =
  [%import: Feedback.feedback]
  [@@deriving yojson]
\end{ocaml}

The above code will generate the following functions:
\begin{ocaml}
val feedback_to_yojson : feedback -> json
val feedback_of_yojson : json -> (feedback, string) Result.result
\end{ocaml}
implementing the bridge between OCaml values and their JSON
representation.

We can use this method to export any data type exposed in the OCaml
plugin API of \coq. As the \jscoq system matures, support for querying
more \coq objects is gradually introduced. For example, using this
mechanism, we can reliably recognize and handle user-specified
notations, printing them in an enhanced way if desired; this is hard
to do without serialization support due to the highly flexible \coq
pretty-printing engine.

\subsection{Document Manager}
\label{sec:coq-manager}

The document manager relates the HTML document model shown to the user
to the internal document maintained by the \coq proof engine. Thanks to
careful API design, the UI state is minimized and the relation can be
mostly implemented in a stateless, reactive way. The manager currently
has three distinct components:
\begin{itemize}
\item a \lstinline!CoqPanel! object, providing the user interface for
  the goal and query buffers;
\item a \lstinline!CoqProvider! abstract object that encapsulates the
  management of \coq statements and which, in particular, takes care of
  selecting the next statement, highlighting, change notifications, etc.;
\item a \lstinline!CoqManager! object that queries the providers and
  coordinates them with the panel and \coq itself, propagating errors
  and logs, and keeping track of the proper \coq state.
\end{itemize}

A key feature of our approach is the use of \href{https://codemirror.net/}{CodeMirror}~\cite{codemirror} as an instance
for the \lstinline!CoqProvider! component. CodeMirror is an open-source software described as ``a versatile text editor implemented in JavaScript for the browser. It is specialized for editing code, and comes with a number of language modes and add-ons that implement more advanced editing functionality.'' In practice, our \coq CodeMirror mode is able to parse and recognize \coq statements, and will notify the manager when a particular part is invalidated by the user.

\subsection{Package Manager}
\label{sec:packaging-system}

The package manager takes care of loading the needed packages and \coq
~\texttt{.vo} files into the \coq instance. Packages are described as JSON
files, stating their dependencies and the set of \coq logical paths the
package provides. Unqualified modules are considered as deprecated by \jscoq. 

The basic unit of the \jscoq package format is the logical path
\lstinline|pkg_id|, which consists of the module identifier, a list of
strings. We assume logical and physical paths equal, thus logical path
\lstinline!A.B.C! will correspond to \lstinline!A/B/C/!. For each
logical path, the package format requires a list of \texttt{.vo} files (the
\lstinline!vo_files! field) and \texttt{.cma} files. For example:
\begin{jslisting}
{
 "pkg_id": [ "Coq", "extraction" ],
 "vo_files": [
   "ExtrHaskellNatInt.vo", "ExtrOcamlString.vo", "ExtrHaskellBasic.vo",
   "ExtrOcamlIntConv.vo", "ExtrHaskellZNum.vo", "ExtrOcamlNatBigInt.vo",
   "ExtrOcamlBigIntConv.vo", "ExtrHaskellZInteger.vo",
   "ExtrOcamlNatInt.vo", "ExtrHaskellNatNum.vo", "ExtrHaskellString.vo",
   "ExtrOcamlZInt.vo", "ExtrHaskellZInt.vo", "ExtrOcamlBasic.vo",
   "ExtrOcamlZBigInt.vo", "ExtrHaskellNatInteger.vo"
  ],
 "cma_files": [ "extraction_plugin.cma" ]
}
\end{jslisting}

The second unit of the \jscoq package manager is the \emph{bundle}. A
bundle is a set of loadpaths, together with a list of dependencies:
\begin{jslisting}
{
  "desc": "math-comp",
  "deps": [],
  "pkgs": [...]
}
\end{jslisting}
Then, the package manager will proceed to load bundles in the
background and notify the IDE when they are ready. We currently
support 16 popular \coq packages, including the full Mathematical
Components library~\cite{gonthier:inria-00258384}.

Adding a new package to \jscoq is reasonably straightforward; the
current process basically consists in adding a few lines to a makefile
pointing where the sources to the package are. Then, our build system
will compile and generate the corresponding JSON file
automatically. Unfortunately, packages must be compiled with the same
exact \coq and OCaml versions used to build \jscoq; thus remote
distribution of packages seems hard for the moment.

\subsection{Document Generation}
\label{sec:document-generation}

Manually writing HTML documents with \coq proofs can be tedious in
some cases. To alleviate this task, we provide \lstinline!udoc!, a
fork of the \coqdoc tool that generates \jscoq documents. For the time
being, the tool is aimed at achieving maximal compatibility with
existing developments using \coqdoc, and the mapping from \coq files
to \jscoq documents is reasonably straightforward.
The \lstinline!udoc! tool has been used to generate the \jscoq versions
of Software Foundations~\cite{Pierce:SF}, Certified Programming with
Dependent Types~\cite{chlipalacpdt2011}, the MTAC
tutorial~\cite{DBLP:conf/icfp/ZilianiDKNV13}, etc.

\section{Practical Validation}
\label{sec:practical-validation}

So far our \jscoq system has behaved in a stable way, filling a niche for
casual \coq users, tutorials, and people willing to try \coq add-ons
in an easy way. Indeed, even at this early stage, \jscoq has already
been used to support some \coq courses and a tutorial, including:
\begin{compactitem}
\item \href{https://team.inria.fr/marelle/en/advanced-coq-winter-school-2016/}{``Winter School Advanced Software Verification and Computer
  Proof''}, Sophia Antipolis, January 18-22, 2016;
\item \href{https://github.com/math-comp/wiki/wiki/tutorial-itp2016}{``Mathematical Components, an Introduction''} and ITP tutorial,
  Nancy, August 27, 2016; and
\item the \href{https://math-comp.github.io/mcb/}{``Mathematical
    Components'' book}.
\end{compactitem}
These two courses total around one hundred \jscoq users. Feedback was
quite positive, as the tool worked well for everybody and allowed them
to experience \coq; the main complaints came from the immaturity of the
user interface, in particular the fact that our choice of UI panels does not adapt
well to different screen sizes.

Regarding the \jscoq instance hosted in our servers, it would be safe
to say that more than a thousand unique users have tried to access it.
However, note that we do not track or gather any personally
identifiable information, and so this estimate has been done on traffic data only and
may not be very exact. There is also interest in using \jscoq to teach
established \coq classes; work is underway and we believe this could
happen soon, as the tool approaches its 1.0 release.
Additionally, the tool has been proved valuable in question and
answers sites such as \url{http://stackoverflow.com}, where it allowed
users to share runnable \coq code snippets.

Performance-wise, our experimental findings are consistent with the predictions by
the \jsoo developers: \jscoq usually runs on par or faster than the
bytecode version of \coq. Memory use is acceptable, with a \jscoq
instance having loaded the \coq prelude plus the Mathematical
Components library topping at 300MiB in Chrome 54.0. We have also
found that the stability of a particular \coq release is quite
consistent; thus, releases that tend to run well usually continue to
do so in future browser versions. Note however that newer \coq versions
may introduce performance problems by themselves. This is usually
difficult to predict, as, of today, \coq upstream changes are not
tested in our backend before integration occurs.

\section{Related Work}
\label{sec:related-work}

There exist a large amount of work in the domain of hybrid document
systems, starting with the WEB system~\cite{knuth1984literate},
mathematical software~\cite{Maple10,ram2010}, and web-based
solutions~\cite{PER-GRA:2007,eloquent-js,edukera,mathbox}.

In the realm of theorem proving, we can highlight Proof
General~\cite{DBLP:conf/tacas/Aspinall00} as a popular tool for Emacs
users, providing a very comprehensive feature set and significantly
easing script editing for several interactive proof assistants. Along
with CoqIDE~\cite{Coq:manual}, this is the standard choice for \coq
users. Proof General communicates with \coq using a non-structured,
text-based protocol, but it has been recently updated to support the
more modern XML protocol~\cite{pg-xml}.
Advanced coding features such as completion and improved display are
provided by the Company Coq~\cite{CompanyCoq2016} Proof General
add-on. The use cases provided by Company Coq have been very useful in
shaping the direction of our own system.

The idea of
proof-by-pointing~\cite{DBLP:journals/fac/Bertot99,DBLP:conf/tacs/BertotKT94}
--- which allows the user to develop proofs by interacting with
logical connectives --- also provided guidance and motivation for the
development of our tool.

We are obviously indebted to the work of Barras et
al~\cite{DBLP:conf/itp/BarrasTT15}, which introduced most of the
\coq-level technology allowing the development of \jscoq, work which
is itself inspired by developments in the Isabelle/jEdit editing
technology~
\cite{DBLP:conf/itp/Wenzel14,DBLP:journals/corr/abs-1304-6626}.
Other interesting IDE efforts for the \coq system
are~\cite{Faithfull_2016,DBLP:journals/corr/abs-1304-6626,bengtson2013kopitiam}.

To the best of our knowledge, \jscoq is one of the first systems to
embed a full theorem prover inside a browser. A prominent web-based
system that depends on a server is ProofWeb~\cite{Kaliszyk200749},
which provides a web interface to a \coq server and many other theorem
provers. The Logitext~\cite{logitext} approach is related to
proof-by-pointing, but web-based and specialized for sequent calculus
proofs.

PeaCoq~\cite{peacoq} is a web-based front end for \coq. The \jscoq
editor component was derived from its editor
implementation. Currently, PeaCoq provides much richer proof-building
capabilities than \jscoq, but it relies on a central server. 
This situation has recently changed, as the PeaCoq development version
is based on SerAPI~\cite{gallegoarias:hal-01384408}. Thus, we hope for
the two tools to share the same back end and protocol in the near
future. Given that SerAPI is capable of running as a Web Worker, this
would also free PeaCoq from its central server dependency, if so desired.

\section{Future Work}
\label{sec:future-work}

In its current iteration, we consider the \jscoq prototype to be a usable
beta; however, quite a bit of work still remains to be done. The
pending tasks can be categorized into three main areas:
\begin{compactitem}
\item user interface and web platform;
\item document generation;
\item protocol and technical foundations.
\end{compactitem}
We provide below more details on each specific point.
\paragraph{User interface and web platform}
Currently, we are taking little advantage of the possibilities of the
web platform. Apart from a few experiments, the text-based
roots and workflow of \coq are still much present in \jscoq, as it
builds upon the existing facilities.
Interesting features such as support for mouse interaction or active
document components may be possible to support without big
modifications to \coq. However, some other features will require
upstream changes in order to achieve a robust implementation.

A key challenge is to improve printing; in particular, an
often-requested feature is nice support for mathematical notations. So
far, only some experiments have been completed, but we are optimistic
that real progress will happen here.

Our user interface proper would also benefit from a rework to allow
for more flexibility and functionality.
Some interesting ideas are to provide a more data-centric view of the
goal buffer, with modifications and goal history.
A key challenge is to choose an interesting ``web framework'' upon
which to base our work. So far, the \emph{phosporjs} framework,
developed for the Jupyter scientific programming community, looks very promising.
Also, our CodeMirror mode needs some additional work.

Regarding our platform support, we are missing two important features:
support for \coq's virtual machine (VM) and support for reliable
timeout/interruptions. The VM of \coq is written in C, and is thus
outside of the scope of \jsoo; however, we believe that supporting it
would be possible by using the emscripten~\cite{emscripten}
transpiler. This task is however low-priority as our users will rarely
choose the web platform for high performance computing. The second
point is more delicate, as both timeout and interruptions are
implemented in \coq by using Unix signals. Indeed, both Microsoft Windows and
the browser environment lack proper signal support, and, while some
workarounds exist, they are still far from complete. In the near
future, we plan to use such workarounds; however, in some cases, killing
the whole \jscoq process is required if the user wants to interrupt a
long-running computation.

\paragraph{Document generation}
Given our intended audience, having good content-generation tools is
very important. In general, all the challenges of book and web page generation do
apply to our current interactive document creation workflow.
There are already plans for extending the current \lstinline!udoc! tool or incorporating \jscoq
support to other document generators, and we would like to coordinate
with the rest of the \coq ecosystem to achieve smooth integration. In
particular, it would be very useful to have good support for
exercise-style boxes, as well as allowing for extended interaction of the
student with particular \coq features (think, for instance, ``click to show
the type'', or ``hover to show definition'' on selected terms).
Longer term, we would like to provide a better
document-generating experience than the one of the CoqDoc/\lstinline!udoc! tool. %

Also, support for interactive elements --- think of a ``\coq box'', i.e., an
HTML \texttt{div} that is filled with the result of the execution of some \coq
program --- is a must. We are also very close to providing
Proviola-like~\cite{Tankink_2010} functionality, that is to say, a
system where the user can interactively perform proof replay.

An interesting, but insufficiently developed, initiative is our
``CollaCoq''\footnote{\url{https://x80.org/collacoq/}} service. At CollaCoq,
users can paste arbitrary Coq programs, obtaining a link that they can
share, so others can run and experiment with the script. While
functional, the service would benefit from additional work.

\paragraph{Protocol and technical foundations}
As we have previously discussed, work on \jscoq has spurred the
creation of a new communication protocol with \coq. Dubbed
\emph{SerAPI}~\cite{gallegoarias:hal-01384408}, for ``Seralization-based API'', this protocol provides a
more data-centric view of the \coq system and will be at the core of future \jscoq versions.
A full discussion of SerAPI is outside the scope of this paper,
but we believe that it can get \coq closer to interesting scientific
computing initiatives such as the OpenDreamKit project\footnote{\url{http://opendreamkit.org}}, which intends to provide an open toolkit for the development of mathematics-focused Virtual Research Environments.

\section{Conclusion}
\label{sec:conclusion}

We have introduced and detailed \jscoq, a new web-based execution
platform and user environment for the Coq interactive proof assistant.
\jscoq is just a set of static HTML/JavaScript files; thus it can be
easily hosted by anyone, and used locally or offline.

Even though \jscoq is still a work in progress, we believe that the
possibilities are exciting, and we think it is not too presumptuous to
state that this experiment has brought some fresh air to the area of
\coq user interfaces. The tool already enjoys some practical impact:
it has been used to give some courses, with more courses and tutorials
planned in the near future.

\section*{Acknowledgments}

We would like to thank Clément Pit--Claudel and Enrico Tassi for
discussion over this work, the developers of \jsoo  for their support, and
the anonymous reviewers for their insightful comments and careful
reading of our paper. This research has been funded by the ANR FEEVER
project.

\bibliographystyle{eptcs}
\bibliography{jscoq}

\end{document}